\documentstyle[12pt,epsfig]{article}
\def\be{\begin{equation}} \def\ee{\end{equation}} \def\bea{\begin{eqnarray}}
\def\eea{\end{eqnarray}} \def\nnb{\nonumber}

\begin{document}


\hfill{\tt USC(NT)-Report-00-4}

\hfill{October 6, 2000}

\begin{center}
\vskip 1.0cm {\Large\bf
Capture rate and neutron helicity asymmetry for ordinary muon 
capture on hydrogen 
}
\vskip 1.cm {\large 
Shung-ichi Ando\footnote{E-mail:sando@nuc003.psc.sc.edu}, 
Fred Myhrer\footnote{E-mail:myhrer@sc.edu}, 
and Kuniharu Kubodera\footnote{E-mail:kubodera@sc.edu}
}\\ 
\vskip 1.cm {\large \it  
Department of Physics and Astronomy,
University of South Carolina, Columbia, SC 29208
}\\ 
\vskip 1cm
\end{center}

Applying heavy-baryon chiral perturbation theory
to ordinary muon capture (OMC) on a proton,
we calculate the capture rate and 
neutron helicity asymmetry
up to next-to-next-to-leading order.
For the singlet hyperfine state,
we obtain the capture rate
$\Gamma_0 =695$ sec$^{-1}$ 
while, for the triplet hyperfine state,
we obtain the capture rate 
$\Gamma_1=11.9 $ sec$^{-1}$
and the neutron asymmetry $\alpha_1=0.93$.
If the existing formalism is used 
to relate these atomic capture rates
to $\Gamma_{\rm liq}$, the OMC rate in liquid hydrogen, 
then $\Gamma_{\rm liq}$ corresponding 
to our improved values of  
$\Gamma_0$ and $\Gamma_1$ is found to be 
significantly larger than the experimental value,
primarily due to the updated larger value of $g_A$.
We argue that this apparent difficulty may be correlated
to the specious anomaly recently reported 
for $\mu^-+p\rightarrow n+\nu_\mu+\gamma$,
and we suggest a possibility 
to remove these two ``problems" simply and simultaneously
by reexamining the molecular physics input
that underlies the conventional analysis of $\Gamma_{\rm liq}$.


\newpage

\noindent {\bf 1. Introduction}

Muon capture on a proton is a valuable source of information
about $g_{\mbox\tiny{P}}$,
the pseudoscalar coupling constant
of the nucleon weak current \cite{primakoff}.
One can study two processes, ordinary muon capture (OMC)
and radiative muon capture (RMC):\footnote{
In this article OMC and RMC always refer to capture
in a hydrogen target.}
\bea
&&\mu^- + p \to \nu_\mu + n  \ \ \ \, \;\;\;\;\;\;\;({\rm OMC}) ,
\label{eq;omc}
\\
&& \mu^- + p \to \nu_\mu + n +\gamma\;\;\;\;\;({\rm RMC}).
\label{eq;rmc}
\eea
Heavy-baryon chiral perturbation theory (HBChPT)
is well suited for describing these processes,
which involve small enough energy-momentum transfers
to render the HBChPT series rapidly convergent.
We report here our HBChPT calculation for OMC 
up to next-to-next-to-leading order.
To explain the motivation and significance 
of our work, we first describe briefly
the current status of OMC and RMC.

The OMC rate in liquid hydrogen,
$\Gamma_{\rm liq}$, has been measured
with 5\% accuracy \cite{bardin81}:
\bea
\Gamma_{\rm liq}^{exp}
=460\pm 20\ \ \ \mbox{\rm [sec$^{-1}$]}.
\label{eq:Gamma-liq}
\eea
Theoretically, one first calculates the atomic OMC rates 
for a proton, 
$\Gamma_0$ and $\Gamma_1$, where
the suffix ``0" (``1") refers to the singlet (triplet)
hyperfine state of the hydrogen atom.
To compare with experiment,
one needs a theoretical framework 
to relate $\Gamma_0$ and $\Gamma_1$
to $\Gamma_{\rm liq}$.
For convenience, we refer to this framework
as the ``atom-liquid" translation formulae.
A great deal of experimental and theoretical effort
has been invested on these translation formulae \cite{bakalov}.
In Refs. \cite{primakoff,opat},
$\Gamma_0$ and $\Gamma_1$ were calculated 
using the phenomenologically parametrized
weak nucleon form factors with the PCAC value for $g_P$.
If we combine these estimates with 
the existing atom-liquid translation formulae,
the resulting $\Gamma_{\rm liq}^{theor}$
agrees with $\Gamma_{\rm liq}^{exp}$
within the experimental error \cite{bakalov}.
In HBChPT, $\Gamma_0$ and $\Gamma_1$
were calculated up to next-to-leading order (NLO)
by Bernard {\it et al.} \cite{bernard-omc-rmc};
it was also reported that a one-loop level 
calculation reproduces the known analytical PCAC correction
for $g_P$, see Refs. \cite{bernardetal94,FS}.
On the other hand, 
the precise empirical determination of $g_P$
is hampered by the 5\% error in $\Gamma_{\rm liq}^{exp}$;
the sensitivity of $\Gamma_{\rm liq}^{exp}$
to $g_P$ is rather modest because the fixed momentum transfer
in OMC ($q^2=-0.88m_\mu^2$) 
suppresses the contribution of the $g_P$ term, 
which contains the pion-pole structure 
$\sim\frac{1}{q^2-m_\pi^2}$.

Since RMC is free from this kinematic constraint,
it can be a more sensitive probe of $g_P$,
despite its extremely small branching ratio.
A recent TRIUMF experiment succeeded in measuring 
$d\Gamma_{\rm RMC}/dE_\gamma$,
the absolute photon spectrum for RMC in liquid hydrogen
\cite{jonkmans96,wright98}.  
If one uses atomic RMC amplitudes calculated
in the phenomenological minimal coupling method
\cite{fea80},
and if one adopts the existing atom-liquid translation formulae,
then the observed $d\Gamma_{\rm RMC}/dE_\gamma$
cannot be reproduced unless $g_P$ 
is artificially increased from its PCAC value
by as much as 50 \%.
However, such a large deviation of $g_P$ from 
its PCAC value is extremely unlikely 
according to an HBChPT calculation \cite{bernardetal94}. 
This very astonishing feature 
reported for $d\Gamma_{\rm RMC}/dE_\gamma$ 
motivated reexamination of the formalism
used in \cite{fea80} to calculate the RMC amplitude, 
and several calculations 
based on HBChPT have been carried out 
\cite{bernard-omc-rmc,mkm98,AM}. 
A next-to-leading order (NLO) calculation \cite{mkm98}
indicates that HBChPT essentially 
reproduces $d\Gamma_{\rm RMC}/dE_\gamma$
given in \cite{fea80}.
A next-to-next-to-leading order (NNLO)
calculation \cite{AM} has confirmed 
that the HBChPT expansion converges rapidly 
and that loop corrections to 
$d\Gamma_{\rm RMC}/dE_\gamma$ are tiny. 
Furthermore, a recent calculation \cite{bernard-omc-rmc}
that incorporates the explicit $\Delta$ degrees of
freedom into a tree-diagram HBChPT calculation 
suggests that the inclusion of the $\Delta$
modifies the spectrum only by 5\%, 
a result consistent with the earlier finding 
of Beder and Fearing \cite{bf89}. 
This change is not large enough to remove 
the above-mentioned anomalous $g_P$ value.
Thus the systematic analyses based on HBChPT
strongly indicate that no drastic changes
in the atomic RMC amplitudes from the existing
estimates should be expected.
It then seems likely that the 
$d\Gamma_{\rm RMC}/dE_\gamma$ problem
is caused by the currently adopted 
atom-liquid translation formulae.

Meanwhile, an experiment that uses 
a hydrogen {\it gas} target to directly measure 
$\Gamma_0$ with 1\% accuracy is planned at PSI 
\cite{PSIexp}.
The use of the gas target eliminates ambiguities 
due to the molecular capture processes.
The envisaged 1\% accuracy will 
significantly increase precision 
with which the empirical value of $g_P$ is determined.
We note, however, that, to make comparison 
between theory and experiment at the 1\% level, 
the existing estimate of $\Gamma_0$
based on an HBChPT calculation up to NLO
needs to be improved.
First, one must ascertain that 
the input physical constants like $f_\pi$ and $g_A$
have sufficient precision.
Secondly, NNLO loop corrections need to 
be evaluated\footnote{
The transition amplitude of OMC was calculated in HBChPT 
up to NNLO by Fearing {\it et al.} \cite{fearing-omc},
but the capture rate was not explicitly given in that work.
}. 

In this article we present an HBChPT calculation 
for OMC up to NNLO in which the influence of 
uncertainties in the low-energy constants 
is carefully examined.
In addition to $\Gamma_0$ and $\Gamma_1$,
we calculate the neutron helicity asymmetry
(to be defined later).
It is found that $\Gamma_0$ obtained here is
significantly larger than the previous estimates
\cite{primakoff,opat},
whereas $\Gamma_1$ essentially agrees with
the literature values.
The larger value of $\Gamma_0$ is primarily
due to the updated larger value of $g_A$.
A second main point of our paper is to discuss 
the observational ramifications of our new estimates.
If we use the new values 
of $\Gamma_0$ and $\Gamma_1$ 
together with the ``standard" 
atom-liquid translation formulae \cite{bakalov},
the resulting $\Gamma_{\rm liq}^{\rm theor}$ 
turns out to be  significantly larger 
than $\Gamma_{\rm liq}^{exp}$ in Eq. (\ref{eq:Gamma-liq}),
another possible indication that 
the existing atom-liquid translation formulae 
may require reexamination.
We shall argue that this difficulty is probably
related to the above-mentioned ``anomaly" in RMC,
and that there is possibility to resolve these two 
problems simply and simultaneously 
by invoking a molecular mixing parameter
discussed by Weinberg \cite{wein60}.

\vspace{5mm}

\noindent {\bf 2. Heavy-baryon chiral perturbation theory (HBChPT)}

HBChPT is a low energy effective field theory of QCD, 
which has a systematic perturbative expansion 
in powers of $Q/\Lambda_\chi$, 
where $Q$ is a small typical four-momentum scale 
characterizing a process in question,
or the pion mass $m_\pi$;
$\Lambda_\chi$ is the chiral scale,
 $\Lambda_\chi\simeq 4\pi f_\pi
\sim m_N \simeq 1$ GeV. 
A typical scale $Q$ in muon capture (both OMC and RMC)
is the muon mass $m_\mu=105.7$ MeV, 
and hence $Q/\Lambda_\chi\sim 0.1$.
One therefore expects a rapid convergence 
of relevant chiral perturbation series for muon capture;
the explicit HBChPT calculations 
\cite{FS,AM,bernard-JMOD,fearing-omc,fearing-omc-rmc}
are consistent with this expectation.

The effective chiral Lagrangian is expanded as
\bea
{\cal L} = \sum {\cal L}_{\bar{\nu}}
={\cal L}_0+{\cal L}_1+{\cal L}_2+\cdots .
\eea
The subscript $\bar{\nu}$ denotes the order of terms,
$\bar{\nu}=d+n/2-2$, where $n$ is the number of nucleon lines 
and $d$ the number of derivatives or powers 
of $m_\pi$ involved in a vertex.
The terms relevant to our calculation are 
\bea
{\cal L}_0 &=& \bar{N}
\left[iv\cdot D+2ig_AS\cdot \Delta\right]N
+f_\pi^2{\rm Tr}\left(-\Delta\cdot\Delta+
\frac{\chi_+}{4}\right),
\\
{\cal L}_1 &=& \frac{1}{2m_N}\bar{N}\left[(v\cdot D)^2-D^2+
2g_A\{v\cdot \Delta,S\cdot D\}
-i(1+b_5)[S^\alpha,S^\beta]f^+_{\alpha\beta}
\right]N ,
\\
{\cal L}_2 &=& \frac{1}{(4\pi f_\pi)^2}\bar{N}\left[
c_3v^\alpha [D^\beta,f^+_{\alpha\beta}]
+c_{13}g_AS^\alpha[D^\beta,f^-_{\alpha\beta}]
+ic_{14}g_AS^\alpha[D_\alpha,\chi_-]
\right] N+{\cal L}_{1/m_N^2} ,
\eea
where ${\cal L}_{1/m_N^2}$ is the $1/m_N^2$ Lagrangian
given in Ref. \cite{AM}. Furthermore
\bea
D_\mu &=& \partial_\mu+\Gamma_\mu,
\\
\Gamma_\mu &=& 
\frac12[\xi^\dagger,\partial_\mu\xi]
-\frac{i}{2}\xi^\dagger F^R_\mu\xi
-\frac{i}{2}\xi F_\mu^L\xi^\dagger,
\ \ \ 
F^{R,L}_\mu = \frac{\vec{\tau}}{2}
\cdot(\vec{v}_\mu\pm\vec{a}_\mu),
\\
\Delta_\mu &=&
\frac12\{\xi^\dagger,\partial_\mu\xi\}
-\frac{i}{2}\xi^\dagger F^R_\mu\xi
+\frac{i}{2}\xi F_\mu^L\xi^\dagger,
\\
f^\pm_{\mu\nu} &=& \frac12\left(
\xi^\dagger F^R_{\mu\nu}\xi \pm\xi 
F^L_{\mu\nu}\xi^\dagger\right),
\ \ \ 
F^{R,L}_{\mu\nu} =\partial_\mu 
F^{R,L}_\nu-\partial_\nu F^{R,L}_\mu
-i[F^{R,L}_\mu,F^{R,L}_\nu] ,
\\
\chi^\pm &=& 
\xi^\dagger\chi\xi^\dagger\pm\xi\chi^\dagger\xi,
\ \ \ \chi=m_\pi^2, 
\ \ \ \xi={\rm exp}(i\vec{\tau}\cdot\vec{\pi}/2f_\pi). 
\eea 
In these expressions 
$\vec{v}_\mu$ and $\vec{a}_\mu$ are the  
isovector vector and axial-vector
external fields, respectively;
$v^\mu=(1,\vec{0})$ is the velocity four-vector,
and $S^\mu=(0,\vec{\sigma}/2)$ is the nucleon spin operator.
We ignore the isospin breaking effect
and use $m_N=(m_p+m_n)/2$ as the nucleon mass\footnote{
In calculating the transition amplitudes
we ignore isospin breaking but, in evaluating
the phase space, we do retain the neutron-proton
mass difference (see later).
\label{footnote;breaking}}. 
Our effective Lagrangian contains
the {\it low energy constants} (LECs),
$b_5$, $c_3$, $c_{13}$ and $c_{14}$.\footnote{
Our notations for the LECs are different 
from those in Ref. \cite{bernard-JMOD}.
The relations between them are
\bea
b_5= c_6,\ \ \ c_3= B_{10},\ \ \ 
c_{13}=2(4\pi f_\pi)^2B_{24}/g_A,\ \ \ c_{14}=B_{23}. 
\nonumber
\eea}
The LECs, $b_5$, $c_{13}$ and $c_{14}$, are 
finite constants fixed by experiments. 
To one-loop order, 
$b_5$ = $\kappa_V + g_A^2m_\pi m_N/4\pi f_\pi^2$, 
where
$\kappa_V$ =3.706 is the isovector anomalous magnetic moment. 
The constant $c_{13}$ is fixed 
by the mean square axial radius deduced from 
(anti) neutrino-proton scattering, 
$\langle r_A\rangle=0.65$ fm \cite{rA};
its numerical value is $c_{13}$ = $(4\pi f_\pi)^2 
\frac{\langle r_A\rangle^2}{3}$ = 4.88.  
The parameter $c_{14}$ is fixed by the GT discrepancy 
defined by
\bea
\Delta_{GT} \equiv 1-\frac{g_Am_N}{f_\pi g_{\pi N}}
=- \frac{2m_\pi^2}{(4\pi f_\pi)^2}c_{14} ,
\label{eq;GTdelta}
\eea
where $g_{\pi N}$ is the $\pi NN$ coupling constant.
The one-loop diagrams are renormalized by $c_3$. 
Integrating in $d=4-2\epsilon$ dimension, we have
\bea
c_3 = -\frac16(1+5g_A^2)R+c_3^{R},
\ \ \ {\rm with} \ \ \ 
R= \frac{1}{\epsilon}+1-\gamma+ 
{\rm ln}\frac{4\pi\mu^2}{m_\pi^2} ,
\eea
where $\gamma=0.5772\cdots$, and the mass scale $\mu$ is 
a parameter in dimensional regularization. 
Note that we include $\mu$ into $c_3^{R}$
to avoid the $\mu$ dependence in the amplitudes. 
The parameter $c_3^R$ is fixed by the empirical radius 
of the isovector Dirac form factor 
$\langle r_1^V\rangle^2=0.585$ fm$^2$ \cite{rV}. 
Thus, from 
$\frac{\langle r_1^V\rangle^2}{6}=
\frac{c_3^R}{(4\pi f_\pi)^2}-
\frac{1+7g_A^2}{96\pi^2 f_\pi^2}, $
we deduce $c_3^R=5.39$ (5.42) with $g_A=1.26$ (1.267). 

\vspace{5mm}

\noindent {\bf 3. Atomic OMC rates and 
neutron helicity asymmetry: Formalism}

The OMC process is effectively described 
by the current-current interaction. 
Thus the transition amplitude reads
\be
M_{fi}=\frac{G_\mu V_{ud}}
{\sqrt{2}}\,L_\alpha J^\alpha,
\ee
where $G_\mu V_{ud}\equiv G_\beta$
is the Fermi constant, 
$L_\alpha$ is the leptonic weak current,
and $J^\alpha$ is the nucleon weak current.
The leptonic current is simply given by
$L_\alpha = 
\bar{u}_\nu\gamma_\alpha(1-\gamma_5)u_\mu$,
whereas $J^\alpha$
is a much more complex object reflecting hadron dynamics.
Here we evaluate $J^\alpha$ in HBChPT
up to NNLO (one-loop) chiral order.

Since in HBChPT the nucleon current $J^\alpha$ 
is expanded in terms of $1/m_N$, 
it is convenient to write $J^\alpha$ in the 
{\it Pauli}-spinor form.
The time and spatial components 
of the nucleon current $J^\alpha=J^\alpha_V-J^\alpha_A$
are written as 
\bea
&& J^0_V(q)=f_1^V(q),\ \ 
\vec{J}_V(q)=i\vec{\sigma}\times 
\hat{q}f_2^V(q)+\hat{q}f_3^V(q),
\label{eq;JV}
\\ &&
J^0_A(q)=\vec{\sigma}\cdot \hat{q} f_3^A(q),\ \ 
\vec{J}_A(q) = \vec{\sigma}f_1^A(q)+
\hat{q}(\vec{\sigma}\cdot \hat{q})f_2^A(q),
\label{eq;JA}
\eea
where we have suppressed the initial- and final-nucleon
spinors as well as the common factor $2m_N$.
The form factors,
$f_i^V$ and $f_i^A$ ($i=1,2,3$), introduced here
may be referred to as the non-relativistic (NR)
form factors.  The relations between
the NR form factors and the {\it standard}
relativistic nucleon weak form factors 
are given in the Appendix.
Note that $f_1^V$ and $f_3^V$ are in fact related 
by current conservation. 

The NR form factors calculated up to one-loop order
in HBChPT read
\bea
f_1^V &=& 1+ \frac{c_3^R}{(4\pi f_\pi)^2}q^2 
-\frac{1+17g_A^2}{18(4\pi f_\pi)^2}q^2 
+ \frac{1}{4m_N^2}\left(-\frac{3}{2}+\kappa_V\right)q^2
\nnb \\ &&
+\frac{1}{(4\pi f_\pi)^2}\left[\frac{2}{3}(1+2g_A^2)m_\pi^2
-\frac{1+5g_A^2}{6}q^2\right]f_0(q),
\\
f_2^V &=& \left[\frac{1}{2m_N}(1+\kappa_V) 
+\frac{g_A^2}{64\pi f_\pi^2 m_\pi}q^2 
+\left(\frac{g_A}{4\pi f_\pi}\right)^2
\frac{\pi(4m_\pi^2-q^2)}{4m_\pi}
m_0(q)\right]|\vec{q}|,
\\
f_3^V &=& \frac{|\vec{q}|}{2m_N},
\\
f_1^A &=& g_A\left[1
+\left(\frac{c_{13}}{2(4\pi f_\pi)^2}-
\frac{1}{8m_N^2}\right)q^2\right],
\\
f_2^A &=& g_A\left[\frac{c_{13}}{2(4\pi f_\pi)^2}
+ \Delta_\pi(q)
\left(1-\frac{2m_\pi^2c_{14}}{(4\pi f_\pi)^2}
+ \frac{q^2}{8m_N^2}\right)\right]|\vec{q}|^2,
\\
f_3^A &=& \frac{g_A}{2m_N}\; 
(1-\Delta_\pi(q)q^2) \; |\vec{q}|,
\eea
where $\Delta_\pi(q)^{-1}= (q^2-m_\pi^2)^{-1}$ is 
the renormalized pion propagator,
and the one-loop functions are given by
\bea
f_0(q) &=& \int^1_0dx\,{\rm ln}[1-x(1-x)q^2/m_\pi^2],
\\
m_0(\vec{q}) &=& 1-\int^1_0dx
\frac{1}{\sqrt{1+x(1-x)\vec{q}^2/m_\pi^2}} \; .
\eea

The total OMC rate from a muonic hydrogen atom 
in a hyperfine state $S$ is given as 
\be
\Gamma_S = K\cdot 
\frac{1}{2S+1}\sum_{S_z,h}|M(h;S,S_z)|^2,
\label{eq;gammaS}
\ee
with 
\be
K\equiv \frac{|\phi_\mu(0)|^2}{16\pi m_\mu m_p}
\left(\frac{E_\nu}{E_\nu+\sqrt{m_n^2+E_\nu^2}}\right).
\label{eq;K}
\ee
The helicity amplitude $M(h;S,S_z)$ 
in Eq. (\ref{eq;gammaS}) is specified
by the final neutron helicity $h$ ($h=L,\,R$),
the initial hyperfine spin $S$, 
and its $z$-component, $S_z$.
In Eq. (\ref{eq;K}), 
$E_\nu$ is the final neutrino energy 
given by\footnote{See footnote \ref{footnote;breaking}.}
\be
E_\nu = \frac{(m_\mu+m_p)^2-m_n^2}
{2(m_\mu+m_p)}= 99.15 \ \ [{\rm MeV}].
\ee
The factor $\phi_\mu(0)$ appearing in $K$ is the value 
at the origin of the radial wavefunction for 
the $\mu^-$-$p$ ground state; thus
$|\phi_\mu(0)|^2=(\alpha m^{r}_\mu)^3/\pi$,
where $\alpha$ is the fine structure constant
and $m_\mu^r=m_\mu m_p/(m_\mu+m_p)$. 

When the neutron helicity is monitored,
we define the neutron helicity asymmetry as  
\bea
\alpha_S=\frac{\Gamma_S(L)-\Gamma_S(R)}
{\Gamma_S(L)+\Gamma_S(R)}\,, 
\label{eq;asym}
\eea
where $\Gamma_S(L)$ ($\Gamma_S(R)$)
is the rate of OMC from an initial 
atomic spin state, $S$, leading to a final left-handed
(right-handed) neutron.\footnote{
Thus, if the neutron helicity is not monitored,
$\Gamma_S = \Gamma_S(L) + \Gamma_S(R)$.}

In calculating the capture rates 
for different neutron helicities, 
it is convenient to choose the direction 
of the emitted neutrino as the $z$-axis.  
In general there are eight helicity amplitudes, 
but with this particular choice
we have only three non-zero amplitudes.
They are
\bea
&& M(L;0,0) = 
\frac{\beta}{\sqrt{2}}(f_1^V+2f_2^V+f_3^V+3f_1^A+f_2^A+f_3^A) ,
\\
&& M(L;1,0) = 
\frac{\beta}{\sqrt{2}}(f_1^V-2f_2^V+f_3^V-f_1^A+f_2^A+f_3^A), 
\\
&& M(R;1,-1) = \beta(f_1^V+f_3^V-f_1^A-f_2^A-f_3^A) ,
\eea
where $\beta \equiv 4G_\beta m_N\sqrt{m_\mu E_\nu}$. 
Correspondingly, we have  
$\Gamma_0(L) = K\, |M(L;0,0)|^2,
\Gamma_0(R)= 0, \Gamma_1(L)= K \frac13 |M(L;1,0)|^2$,
and $\Gamma_1(R)= K\frac13 |M(R;1,-1)|^2$.

\vspace{5mm}

\noindent {\bf 4. Atomic OMC rates and 
neutron helicity asymmetry: numerical results}

As emphasized above, at the level of precision 
of our concern, we need to be particularly careful
about the accuracy of the input physical parameters.
The most updated value of $G_\beta$ is
$G_\beta/\sqrt{2}=
0.8030\pm 0.0008\times 10^{-5}$ GeV$^{-2}$. 
For $g_A$ and $g_{\pi N}$,
we use as the standard values $g_A=1.267$ \cite{PDG} 
and $g_{\pi N}=13.4$.

Since the momentum transfer for OMC is fixed,
we calculate the NR form factors for 
$|\vec{q}|=E_\nu$ and $q^2=m_\mu^2-2m_\mu E_\nu$.
The results for each order of HBChPT expansion
are given in Table \ref{table;FF}.
The table clearly shows  
that the chiral perturbation series 
converges very rapidly.
As for the helicity-dependent amplitudes, we obtain 
\be
M(L;0,0)=3.45\, \beta, \ \
M(L;1,0)=-0.77\, \beta,\ \ M(R;1,-1)=0.15\, \beta.
\ee
\begin{table}
\begin{center}
\begin{tabular}{|c||cccccc|} \hline
    & $f_1^V$ & $f_2^V$ & $f_3^V$ & $f_1^A$ & $f_2^A$ & $f_3^A$ \\ \hline 
LO  & 1.000   &  0      & 0       & 1.267   & $-0.426$& 0       \\
NLO & 0       & 0.248   & 0.053   & 0       & 0       & 0.045   \\
NNLO& $-0.030$& $-0.004$& 0       & $-0.021$& 0.006   & 0       \\ \hline
Total& 0.970  & 0.244   & 0.053   & 1.246   & $-0.419$& 0.045
\\ \hline
\end{tabular}
\caption{Numerical values of the OMC form factors
in Eqs. (\ref{eq;JV}) and (\ref{eq;JA}),
calculated for each chiral order 
with the use of $g_A=1.267$ and $g_{\pi N}=13.4$.}
\label{table;FF}
\end{center}
\end{table}

In Table \ref{table;OMCrate} we give our numerical results 
for the atomic capture rates, $\Gamma_0$ and $\Gamma_1$, 
along with HBCHPT calculations of Bernard {\it et al.}
\cite{bernard-omc-rmc,bhm00b}. 
The second column labeled ``This work" represents
the results obtained in our NNLO-calculation. 
Comparing this column with the third column
that gives the results of our NLO-calculation, 
we note that the NNLO corrections decrease $\Gamma_0$ 
significantly (3.9\%). 
Thus it is clear
that, in order to achieve theoretical precision
that matches the 1\% accuracy of
the planned PSI experiment on $\Gamma_0$ \cite{PSIexp},
one must take into account the NNLO terms.
\begin{table}
\begin{center}
\begin{tabular}{|c|cccccc|} \hline
  & This work & This work&
Bernard {\it et al.}\cite{bhm00b} &
Bernard {\it et al.}\cite{bernard-omc-rmc} &
Primakoff \cite{primakoff} & Opat \cite{opat}  \\ 
  & (NNLO) & (NLO) & (NNLO) & (NLO) & &        \\ \hline
$\Gamma_0$ & 695  & 722  & 687.4   & 711  & 664$\pm$20      & 634  \\
$\Gamma_1$ & 11.9 & 12.2 & 12.9    & 14.0 & 11.9$\pm$0.7    & 13.3 
\\ \hline
\end{tabular}
\caption{Comparison of calculated atomic OMC rates.
$\Gamma_0$ ($\Gamma_1$) is the capture rate [in sec$^{-1}$]
for the initial singlet (triplet) hyperfine state.
The entries for the columns labeled ``This work (NNLO)" 
and ``This work (NLO)" have been obtained 
with the use of $g_A$ = 1.267 and $g_{\pi N}$ = 13.4.}
\label{table;OMCrate}
\end{center}
\end{table}

We now turn to comparison with the other calculations 
quoted in Table \ref{table;OMCrate}. 
The estimates of Primakoff \cite{primakoff} 
and Opat \cite{opat} are based on the phenomenological 
parameterization of the weak nucleon current
with $g_P$ fixed at its PCAC value;
Primakoff retained the relativistic kinematics,
whereas Opat used a non-relativistic expansion of 
the amplitudes in terms of $1/m_N$. 
Bernard {\it et al.}'s results
come from an NLO-calculation \cite{bernard-omc-rmc} 
and an NNLO-calculation \cite{bhm00b}. 
The results of the two NNLO-calculations, 
the second and fourth columns in Table \ref{table;OMCrate}, 
exhibit a 1 \% difference.
Bernard {\it et al.} 
used a different value of $g_A$ ($g_A=1.26$).
They also expanded  the phase space and atomic wavefunction 
in Eq. (\ref{eq;K}) in powers of $m_\mu/m_N$.
For the sake of comparison, 
suppose that, in our NNLO calculation 
we apply expansion of $m_\mu/m_N$ to the phase space, 
while keeping the exact atomic wavefunction,
but ignoring neutron-proton ($n$-$p$) mass difference.
Then $\Gamma_0$ calculated up to $1/m_N^2$ 
would become $\Gamma_0^{1/m_N^2}=708$ sec$^{-1}$, 
which is to be compared with
$\Gamma_0=$ 695 sec$^{-1}$ in Table \ref{table;OMCrate}.
If we expand both phase space and wavefunction,
ignoring $n$-$p$ mass difference,
we obtain $\Gamma^{1/m_N^2}_0=705$ sec$^{-1}$.
The different $1/m_N$ expansion schemes
show only a tiny numerical difference, 
but, the rates in both cases are enhanced by $\sim$ 2 \%
compared with the exact phase space case.
This is due to ignoring the nucleon mass difference; 
the rate is proportional to $E_\nu^2$ 
and $E_\nu$ increases by 1 \% 
when the $n$-$p$ mass difference is ignored.
In the present work we retain the exact final phase space  
and atomic wave-function;\footnote{  
We do not, however, include the finite-nucleon-size effect
on the atomic wave-function;
this effect is known to reduce the capture rate 
by $\sim$0.4 \%, see e.g., 
Appendix 1 in Ref. \cite{govaerts}.}
viz., we apply HBChPT expansion 
only to the transition amplitudes.  
On the other hand, 
the 1 \% difference between the two NNLO-calculations
in Table \ref{table;OMCrate} may stem
from differences in input parameters as well. 

According to Table \ref{table;OMCrate},  
$\Gamma_0$'s calculated in HBChPT  
(our present result and those of Bernard {\it et al.})
have significantly larger values than 
the earlier theoretical estimates. 
This is primarily due to the fact that
the modern HBChPT calculations employ 
an updated value of $g_A$,
which is larger than the older values.
Primakoff \cite{primakoff} used $g_A=1.24$, 
while Opat \cite{opat} used $g_A=1.22$. 
In order to illustrate the sensitivity of our results
to the input physical parameters,
we show in  Table \ref{table;rates} the values of  
$\Gamma_0$ and $\Gamma_1$  
corresponding to different values of $g_A$ and $g_{\pi N}$;
the LEC $c_{14}$, 
which is determined by $g_A$ and $g_{\pi N}$
via Eq. (\ref{eq;GTdelta}), is also listed.
As can be seen from the last four rows in the table,
for a given value of $g_A$, variations in $g_{\pi N}$
causes only minor changes in the capture rates;
even though these variations lead to a difference 
of a factor of $\sim$3 in $c_{14}$,
the corresponding changes in the rates  
in the last four rows 
are modest; 1.3\% for $\Gamma_0$ 
and  2.5\% for $\Gamma_1$.
Thus the most crucial input parameter here
is $g_A$.

Finally we discuss the helicity asymmetry.
Due to the $V-A$ weak interaction the final neutron 
is purely left-handed 
when an initial atomic state is 
in the hyperfine singlet state ($S=0$); thus we have
$\alpha_0 = 1$ as a trivial identity. 
For the initial hyperfine triplet state ($S=1$), 
the final neutron can have both left- and right-handed 
helicity components, and therefore $\alpha_1$ 
can have a non-trivial value;
our calculation gives $\alpha_1=0.925$.
Thus almost all out-going neutrons are polarized left-handedly, 
a result consistent with Weinberg's observation \cite{wein60}. 

\begin{table}
\begin{center}
\begin{tabular}{|cc||c|ccc|}\hline
$g_A$ & $g_{\pi N}$ & $c_{14}$ & $\Gamma_0$ & $\Gamma_1$ & $\alpha_1$ \\ \hline 
1.22   & 13.40       & $-2.59$ & 656       & 11.3         & 0.865  \\
1.24   & 13.40       & $-2.07$ & 672       & 11.6         & 0.893  \\
1.26  & 13.40        & $-$1.54 & 689       & 11.9         & 0.918  \\
1.26  & 13.05        & $-$0.65 & 692       & 11.6         & 0.927  \\
1.267 & 13.40        & $-$1.36 & 695       & 11.9         & 0.925  \\
1.267 & 13.05        & $-$0.47 & 698       & 11.7         & 0.934  \\
\hline
\end{tabular}
\caption{The LEC $c_{14}$, atomic capture rates $\Gamma_0$ 
and $\Gamma_1$ \mbox{\rm [sec$^{-1}$]},
and the neutron helicity asymmetry $\alpha_1$, 
calculated in HBChPT up to NNLO 
for various choices of $g_A$ and $g_{\pi N}$.}
\label{table;rates}
\end{center}
\end{table}  

\vskip 5mm \noindent 
{\bf  5. OMC rates for liquid hydrogen targets}

As mentioned, in order to relate theoretical estimates
of the atomic OMC rates to the capture rate
measured in liquid hydrogen, 
one needs (in our terminology) the ``atom-liquid" 
translation formulae.
We briefly describe here the atomic and molecular physics
input that underlies these formulae. 
A muon stopped in liquid hydrogen  
quickly forms a muonic atom ($\mu$-p) in its Bohr orbit. 
The atomic hyperfine triplet state (S=1) is rapidly transformed 
into the singlet state (S=0);
this hyperfine transition rate is known to be
$\lambda_{10}\simeq 1.7\times 10^{10}$ sec$^{-1}$
\cite{bakalov}, 
indeed a very large value.
A muonic atom and a hydrogen molecule 
collide with each other to form a $p$-$\mu$-$p$ molecule,
predominantly in its ortho state (with the two proton spins
parallel to each other). 
Let $\lambda_{pp\mu}$ be the rate of transition
from the atomic hyperfine singlet state
to the ortho $p$-$\mu$-$p$ molecular state. 
Meanwhile, the ortho $p$-$\mu$-$p$ molecular state
decays to the lower-lying para $p$-$\mu$-$p$ 
molecular state. 
Let $\lambda_{op}$ stand for this decay rate.
Taking into account these atomic and molecular processes,
one relates $\Gamma_{\rm liq}$, the OMC capture rate 
in liquid hydrogen, to the atomic capture rates
($\Gamma_0$ and $\Gamma_1$) 
via the formula \cite{bakalov}
\bea
\Gamma_{\rm liq} =
\frac{\lambda_0}{\lambda_0+\lambda_{pp\mu}}\Gamma_0
 +\frac{\lambda_{pp\mu}}{\lambda_0+\lambda_{pp\mu}}
  \frac{\lambda_0}{\lambda_0+\lambda_{op}}
\left(\Gamma_{om}+\Gamma_{pm}
\frac{\lambda_{op}}{\lambda_0}\right),
 \label{eq;lrate}
\eea
where $\lambda_0$ is the muon decay rate, 
$\lambda_0=0.455\times 10^6$ sec$^{-1}$.
In this equation $\Gamma_{om}$ ($\Gamma_{pm}$)
represents the rate of muon capture from the ortho (para)
$p$-$\mu$-$p$ molecular state.
These rates are usually calculated using the formula,
\bea
\Gamma_{om}=2\gamma_O
\left(\frac34\Gamma_0+\frac14\Gamma_1\right), \ \
\Gamma_{pm}=2\gamma_P
\left(\frac14\Gamma_0+\frac34\Gamma_1\right). \ \
\label{eq;gammaOP}
\eea
The factors $2\gamma_O$ and $2\gamma_P$ account for 
modifications of the muon wavefunction as it changes 
from the atomic Bohr orbit to the $p$-$\mu$-$p$ 
molecular state;
according to Ref. \cite{bakalov},
$2\gamma_O=1.009$ and $2\gamma_P=1.143$. 
The validity of Eq. (\ref{eq;gammaOP})
will be discussed later in the text.

As for $\lambda_{pp\mu}$, 
there are several conflicting experimental results,
see e.g. Ref.\cite{PSIexp}. 
We quote here the lowest and highest reported values:
$\lambda^{exp}_{pp\mu}=1.89\pm 0.20\times 10^6$ sec$^{-1}$,
and $\lambda^{exp}_{pp\mu}=2.75\pm 0.25\times 10^6$ sec$^{-1}$.
The current theoretical estimate
$\lambda^{th}_{pp\mu}=1.8\times 10^6$ sec$^{-1}$
lies near the lower edge of the lowest experimental value,
which is a rather uncomfortable situation.
Furthermore, the current experimental and theoretical values
for $\lambda_{op}$ do not agree with each other:
$\lambda_{op}^{exp}=4.1\pm 1.4\times 10^4$ sec$^{-1}$ \cite{bardin-lambda_op} 
and $\lambda_{op}^{th}=7.1\pm 1.2\times 10^4$ sec$^{-1}$ \cite{bakalov}. 
Thus it seems fair to say that 
the existing ``atom-liquid" formulae are not totally
free from uncertainties 
and that these ambiguities can affect our interpretation 
of the OMC rate in liquid hydrogen.

In the following we use for $\lambda_{pp\mu}$ the value
adopted in Ref. \cite{bakalov}, 
$\lambda_{pp\mu}=2.5\times 10^6$ sec$^{-1}$;
for $\lambda_{op}$, we use two representative values: 
$\lambda_{op}^{th}$ and $\lambda_{op}^{exp}$.
Using the OMC rates of Primakoff in Table \ref{table;OMCrate}
and $\lambda_{op}^{th}$, 
Bakalov {\it et al.} \cite{bakalov} obtained
$\Gamma_{\rm liq}=490\pm 10$ sec$^{-1}$, 
in good agreement with the data, 
$\Gamma_{\rm liq}^{exp}$ = 460 $\pm$ 20 sec$^{-1}$. 
However, if we use the values 
of $\Gamma_0$ and $\Gamma_1$ obtained 
in our NNLO-HBChPT calculation together with $\lambda_{op}^{th}$, 
then Eq. (\ref{eq;lrate}) gives a much larger rate,
$\Gamma_{\rm liq}= 518$  sec$^{-1}$.
If we adopt $\lambda_{op}=\lambda_{op}^{exp}$,
$\Gamma_{liq}$ becomes even larger:
$\Gamma_{liq}=532$ sec$^{-1}$. 
Thus, the use of the updated values of 
$\Gamma_0$ and $\Gamma_1$ as obtained here
in combination with the commonly used 
``atom-liquid translation formulae"
spoils the previously reported {\it good} agreement
between $\Gamma_{\rm liq}^{theor}$
and $\Gamma_{\rm liq}^{exp}$.
So, in addition to the problem of
the RMC photon spectrum discussed in the Introduction,
another serious problem seems to be lurking in the OMC sector.

\vskip 5mm

\noindent {\bf 6. A mixing molecular parameter to fit the
OMC and RMC data }

In view of the fact that these two problems
occur in the experiments involving liquid hydrogen targets, 
it seems of interest and of importance 
to reexamine the reliability of 
the formulae hitherto used in the literature
to relate the atomic capture rates
to $\Gamma_{\rm liq}$.
Although a thorough investigation of this issue 
is beyond the scope of this article, 
we wish to discuss here a particular aspect
of molecular physics input
which seems relevant to the present issue
but so far has not been fully examined. 
Taking up an early observation made by Weinberg \cite{wein60}, 
consider the possibility that in liquid hydrogen two ortho 
molecular $p$-$\mu$-$p$ spin states, 
$S_{p\mu p}$ = 1/2 and 3/2 may be populated.
If this indeed happens,  
$\Gamma_{om}$ in Eq. (\ref{eq;gammaOP})
should be replaced with 
\begin{eqnarray}
\Gamma_{om}' = \xi\, \Gamma_{om}(1/2) 
+ (1-\xi) \Gamma_{om}(3/2),
\label{eq;xi}
\end{eqnarray}
where $\Gamma_{om}(1/2)=
\Gamma_{om}$ of Eq. (\ref{eq;gammaOP}) and 
$\Gamma_{om}(3/2)=2\gamma_O\Gamma_1$. 
According to Weinberg \cite{wein60}, 
the mixing parameter $\xi$ can be in the range of
$0.5\le \xi \le 1$.
Only theoretical estimates of $\xi$ exist,
and its literature value is 
$\xi\simeq 1$ \cite{xi,bakalov};
to our knowledge $\xi$ has never been measured experimentally.
To study the consequences of $\xi\ne 1$,
we have calculated $\Gamma_{\rm liq}$
with the use of 
Eqs. (\ref{eq;lrate}),(\ref{eq;gammaOP}),(\ref{eq;xi}), 
for several values of $\xi$.
For $\Gamma_0$ and $\Gamma_1$ we have used the results of 
our NNLO calculation
while for $\lambda_{op}$ we have considered
the two representative values
discussed above: $\lambda_{op}^{th}$ 
and $\lambda_{op}^{exp}$.
The results are shown in Table \ref{table;Lam_c}.
\begin{table}
\begin{center}
\begin{tabular}{|c|ccccc|} \hline
$\xi$ & 1.0 & 0.95 & 0.90 & 0.85 & 0.80 \\ \hline
$\Gamma_{\rm liq}(\lambda^{th}_{op})$   
& 518 & 499 & 480 & 461 & 442 \\
$\Gamma_{\rm liq}(\lambda_{op}^{exp})$ 
& 532 & 512 & 492 & 472 & 452 \\ \hline
\end{tabular}
\caption{The OMC rate in liquid hydrogen,
$\Gamma_{\rm liq}$ \mbox{\rm [sec$^{-1}$]}, 
calculated for five different values of  
the molecular mixing parameter $\xi$, Eq. (\ref{eq;xi}), 
and for two typical choices (explained in the text)
of $\lambda_{op}$,
the ortho-to-para transition rate
in a $p$-$\mu$-$p$ molecule. 
The second (third) row corresponds to the use   
of $\lambda_{op}^{th}$ = 7.1 $\times 10^4$ sec$^{-1}$ 
($\lambda_{op}^{exp}$ = 4.1 $\times 10^4$ sec$^{-1}$).} 
\label{table;Lam_c}
\end{center}
\end{table}
The table indicates that, if $0.8\le \xi \le 0.9$, 
the theoretical and experimental values 
of $\Gamma_{\rm liq}$ are in good agreement.
\footnote{ 
If one replaces Eq. (\ref{eq;lrate}) with
an older and simpler 
``atom-liquid" formula
of Ref. \cite{bardin-lambda_op},
the OMC data can be fit with $\xi=1$,
and the RMC data with $\xi\simeq 0.95$
within the framework of an NLO-HBChPT calculation, 
see Ref. \cite{bhm00b}.} 

\begin{figure}
\begin{center}
\epsfig{file=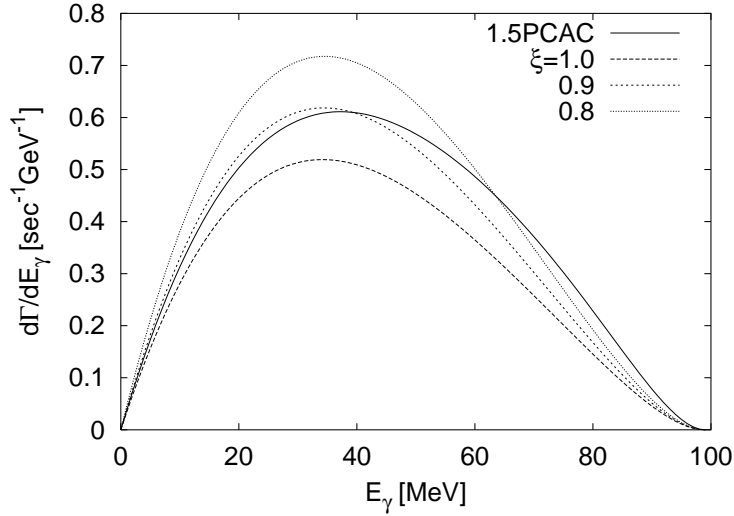,width=10cm}
\caption{The RMC photon spectrum, $d\Gamma_{\rm RMC}/dE_\gamma$. 
The solid line labeled ``1.5PCAC" 
is the result one would obtain from Fearing's model
if the value of $g_P$ is assumed to be
1.5 times the standard PCAC value. 
The other curves represent the results 
of NNLO-HBChPT calculations 
with $\xi=$ 1.0, 0.9 and 0.8.}
\label{fig;omc3-2}
\end{center}
\end{figure}

We next argue that the introduction of $\xi$
in this range leads to resolution 
of the RMC problem as well.
We first remark that
Eqs. (\ref{eq;lrate}),(\ref{eq;gammaOP}),(\ref{eq;xi}),
can be used, mutatis mutandis, for RMC as well,
in particular, for calculation of 
$d\Gamma_{\rm RMC}/dE_\gamma$,
the photon spectrum for RMC in liquid hydrogen.
With the atomic RMC transition amplitudes 
previously obtained
in a NNLO-HBChPT calculation \cite{AM},
we have evaluated $d\Gamma_{\rm RMC}/dE_\gamma$
for various values of $\xi$;
the other atomic and molecular
population parameters are kept fixed
at the values used in Ref. \cite{wright98}:
6.1\% atomic hyperfine singlet state, 
85.4\% ortho $p$-$\mu$-$p$ state
and 8.5\% para  $p$-$\mu$-$p$ state.
The results are shown in Fig. \ref{fig;omc3-2}. 
The dashed line (lowest curve) represents 
the no-mixing case, $\xi$ = 1, 
which corresponds to $d\Gamma_{\rm RMC}/dE_\gamma$
obtained in \cite{AM}.
For comparison , we also show in the figure 
(the solid line)
the result obtained in a modified version 
of the Fearing model \cite{fea80}
wherein $g_P$ is taken to be 1.5 times the PCAC value.
This line represents the best fit curve
to the observed $d\Gamma_{\rm RMC}/dE_\gamma$
in the analysis reported in \cite{jonkmans96,wright98}. 
One can see from the figure that 
$\xi$ in the range of 0.8-0.9 leads to a photon
spectrum that is satisfactorily close to
the ``observed" spectrum (solid line)
for $E_\gamma$ $\geq$ 60 MeV. 

\vskip 5mm

\noindent {\bf  7. Discussion and Conclusions} 

We have considered in our HBChPT calculation
up to NNLO contributions.
Some remarks on possible higher order effects
are in order here.
As discussed in Ref. \cite{bernard-omc-rmc}, 
a one-loop diagram in N$^3$LO 
which contains a vertex with the 
anomalous magnetic moment, $\kappa_V m_\mu/m_N\sim 0.5$,
can be comparable in size to the NNLO diagrams. 
This means that 
the correction to the capture rate 
due to the N$^3$LO terms may reach the 1-2 \% level.
Thus, for a more precise theoretical prediction
of the OMC rates,
it can in principle be important 
to include the N$^3$LO corrections, 
although the existing uncertainty in the 
values of $g_A$ and $g_{\pi N}$ may not warrant the effort.
Moreover, at the level of N$^3$LO, 
the isospin breaking effects 
as well as QED corrections \cite{QEDChPT}
are expected to give sizable contributions.

In conclusion, we have carried out a HBChPT calculation
of the atomic OMC rates to next-to-next-to-leading order.
Our result indicates 
that, once the measurement of the hyperfine-singlet
atomic OMC rate reaches 1\% accuracy, as envisaged
in a PSI experiment,
theoretical predictions based on HBChPT
must include at least NNLO corrections. 
Furthermore, we have shown that 
both the OMC rate and the RMC photon energy spectrum 
measured in liquid hydrogen 
can be reproduced by introducing the molecular 
mixing parameter $\xi$ in the range of $\xi\sim 0.8$ to 0.9. 
It seems interesting to examine whether this range of $\xi$ 
is realistic.  
 
Finally, we have shown that the neutron helicity asymmetry 
for OMC from a hyperfine triplet state is $\sim$93\%.

\vspace{6mm}
\noindent
{\bf Acknowledgment}

The authors are grateful to V\'{e}ronique Bernard, Thomas Hemmert 
and Ulf Mei\ss ner for kindly communicating the content of 
Ref. \cite{bhm00b} and for extremely useful discussions.
This work is supported in part by the NSF 
grant  PHY-9900756.

\vspace{1.5cm}
\noindent
{\bf APPENDIX}

\noindent
Assuming the absence of the second-class current,
one can express the nucleon vector and axial currents
in terms of four form factors:
\bea
J_V^\alpha &=& \bar{u}_n(p')\left[G_V(q)\gamma^\alpha
+ G_M(q)\frac{i\sigma^{\alpha\beta}q_\beta}{2m_N}
\right]u_p(p),
\\
J_A^\alpha &=& \bar{u}_n(p')
\left[G_A(q)\gamma^\alpha\gamma_5
+ G_P(q)\frac{q_\beta}{m_\mu}\gamma_5\right]u_p(p),
\eea
where $G_V(q)$, $G_M(q)$, $G_A(q)$, and $G_P(q)$ are
the vector, weak-magnetism, axial-vector, and pseudo-scalar
form factor, respectively; $q=p'-p$ and 
$u_p$ ($u_n$) is the Dirac spinor for the proton (neutron). 
These standard relativistic form factors are related
to the NR form factors, $f_i^V$'s and $f_i^A$'s,
defined in Eq. (\ref{eq;JA}).
Up to ${\cal O}(1/m_N^2)$
\bea
&&f_1^V = G_V(q)\left(1-\frac{q^2}{8m_N^2}\right)
 + \frac{q^2}{4m_N^2}G_M(q),
\nnb \\
f_2^V &=& \frac{1}{2m_N}\left[G_V(q)+G_M(q)\right]|\vec{q}\; |,
\ \ \ \
f_3^V =  \frac{1}{2m_N}G_V(q)|\vec{q}\; |,
\nnb \\
f_1^A &=& G_A(q)\left(1-\frac{q^2}{8m_N^2}\right),
\ \ \ \
f_2^A = -\frac{G_P(q)}{2m_\mu m_N}\left(1+\frac{q^2}{8m_N^2}\right)
|\vec{q}\; |^2,
\nnb \\
&&f_3^A = \frac{1}{2m_N}
\left(G_A(q)+\frac{G_P(q)q^2}{2m_\mu m_N}\right)
|\vec{q}\; | .
\nnb
\eea

\end{document}